\documentclass[12pt]{article}
\usepackage[dvips]{graphicx}
\usepackage[tight]{subfigure}

\def\be{\begin{equation}}
\def\ee{\end{equation}}
\def\bea{\begin{eqnarray}}
\def\eea{\end{eqnarray}}
\def\beaN{\begin{eqnarray*}}
\def\eeaN{\end{eqnarray*}}
\def\ed{\end{document}}
\def\bit{\begin{itemize}}
\def\eit{\end{itemize}}

\def\sig{\sigma}

\def\lam{\lambda}

\def\Del{\Delta}

\def\Bg{\Bar g}

\def\k{\kappa}
\def\alf{\alpha}

\def\BD{\Bar D}
\def\di{\partial}

\def\Lix{\pounds_\xi}

\def\half{{\textstyle{1 \over 2}}}
\def\~{\tilde}
\def\lag{{\hat{\cal L}}}
\def\m{\label}
\def\l{\left}
\def\r{\right}

\def\goto{\rightarrow}
\def\Bar{\overline}

\def\diag{\rm diag}

\begin{document}

%\large
\centerline{\bf N{\OE}THER AND BELINFANTE CORRECTED TYPES OF
CURRENTS} \centerline{\bf FOR PERTURBATIONS IN THE
EINSTEIN-GAUSS-BONNET GRAVITY}

\smallskip

\centerline{\it { A.N.Petrov}}

\smallskip

\centerline{\it Relativistic Astrophysics group, Sternberg
Astronomical institute,} \centerline {\it
 Universitetskii pr., 13, Moscow, 119992,
RUSSIA}

\centerline{ e-mail: anpetrov@rol.ru}

\smallskip

telephone number: +7(495)7315222

\smallskip

fax: +7(495)9328841

\smallskip


\begin{abstract}
In the framework of an arbitrary $D$-dimensional metric theory,
perturbations are considered on arbitrary backgrounds that are
however solutions of the theory. Conserved currents for
perturbations are presented following two known prescriptions:
canonical N{\oe}ther theorem and Belinfante
symmetrization rule. Using generalized formulae,
currents in the Einstein-Gauss-Bonnet (EGB) gravity for arbitrary
types of perturbations on arbitrary curved backgrounds (not only
vacuum) are constructed in an explicit covariant form. Special
attention is paid to the energy-momentum tensors for perturbations
which are an important part in the structure of the currents.

We use the derived expressions for two applied calculations: a) to
present the energy density for weak flat gravitational waves in
$D$-dimensional EGB gravity; b) to construct the mass flux for the
Maeda-Dadhich-Molina 3D radiating black holes of a Kaluza-Klein type
in 6D EGB gravity.
\end{abstract}

\eject

\section{Introduction}
\m{Introduction}

As multidimensional theories become more and more popular,
the study of the behaviour and properties of solutions in these theories has gained prominence.
In particular, it is especially important to describe perturbations in such theories (see, for example,
\cite{branaPert1} - \cite{branaPert6}, and numerous references there
in). It is very important to construct conserved quantities for such
perturbations. Although several approaches have been developed,
including our recent results (see
\cite{Petrov2009} and references there in),
these have been restricted, as a rule, to only constructing surface ({\em
non-local}) expressions.  The surface expressions
are obtained after integration of so-called superpotentials.
However, keeping in mind cosmological and astrophysical applications
it is important to construct {\em local} conserved quantities (which
usually are expressed by conserved currents for perturbations), and
to connect them with non-local conserved quantities. In this approach,
conservation laws are presented in a form where
superpotentials are connected with correspondent conserved currents.

Arguably, at the present moment, among multidimensional
generalizations of usual 4-dimensional general relativity (4D GR), a
Gauss-Bonnet (GB) modification is the most popular. The action of
the Einstein-Gauss-Bonnet (EGB) theory has a lower (quadratic in
curvature) order of the action of the Lovelock gravity
\cite{Lovelock}. The latter is a generalization of GR, when an
action includes higher order curvature terms preserving the
diffeomorphism invariance and still leading to field equations
containing no more than second order derivatives. On the other hand,
independently, the GB term occurs in the effective lower energy
action of superstring theory \cite{LustTheisen}. The EGB gravity has
many new useful and interesting properties. Therefore in the
framework of the EGB gravity numerous important topics and problems
are intensively studied. They are multidimensional black hole
solutions, black hole thermodynamics and conserved charges, AdS/CFT
correspondence, wormhole solutions and their properties,
cosmological dynamics, membrane paradigm, {\em etc}. It is
impossible to give a full bibliography on studies related to
Lovelock and EGB theories; for a review and further references one
can recommend, e.g., \cite{Charmousis1}, where many aforementioned
aspects are discussed.

In \cite{Petrov2009}, where superpotentials and correspondent
charges in EGB gravity were constructed, we have used the following
three approaches well known in 4D GR. The first approach,
{\em canonical} (direct application of N{\oe}ther
theorem), starts from the Einstein pseudotensor \cite{Einstein16}
and the Freud superpotential \cite{Freud39}. The final and maximally
generalized form in 4D GR is presented by  Katz, Bi\v c\'ak  and
Lynden-Bell \cite{KBL}. The second approach is based on the {\em
Belinfante symmetrization} method \cite{Belinfante}, which firstly
in 4D GR has been applied by Papapetrou \cite{Papapetrou48} for
symmetrization of the Einstein pseudotensor and for the correction
of the Freud superpotential. Maximally generalized application of
the Beinfante method in 4D GR is presented in the works
\cite{PK-lett, PK}. The third approach is frequently called as {\em
field-theoretical} (or {\em symmetrical}) one, where all
perturbations (including metric ones) are presented as an united
field configuration, which propagates in a background spacetime and
is described by symmetrical (metric) energy-momentum tensor. For
review of the above methods see \cite{Petrov2008}.

To the best of our knowledge, unlike superpotentials, authors have
not paid attention to constructing currents in modified theories.
Here, at least in part, we try to close this gap. Following
the proposals in \cite{Petrov2008}, we present currents of a
generalized form in an arbitrary metric theory in the {\em
canonical} and {\em Belinfante symmetrization} approaches. Following
this, the generalized formulae are used to construct the currents in
EGB gravity. Thus, continuing the research begun in \cite{PK2003a}
and \cite{Petrov2009}, we add superpotentials presented in
\cite{Petrov2009} by corresponding currents. The symmetrical
approach, due to its technical particularities, requires a separate
investigation, therefore we do not consider it here.

The paper is organized as follows. In section \ref{D-dimensions},
general definitions in an arbitrary $D$-dimensional metric theory
are given, and general identities necessary for constructing
conserved quantities are presented and discussed.  In section
\ref{CurrentsInD-dimensions}, in an arbitrary metric theory
conserved currents for arbitrary perturbations on arbitrary curved
backgrounds that are however solutions of the theory, are
presented in the framework of both the approaches. In section
\ref{EGBcurrents}, the results of section
\ref{CurrentsInD-dimensions} are used to construct explicit
covariant expressions for the currents in EGB gravity. In section
\ref{Applications}, the new expressions for the currents are used
to construct energy density for weak flat gravitational waves and
mass flux for radiating solutions in the EGB gravity. The concluding remarks
are placed in the last section. In Appendix, we present the necessary, although somewhat
cumbersome, expressions from EGB gravity.

\section{Arbitrary $D$-dimensional
metric theories. The main identities}
 \m{D-dimensions}
 \setcounter{equation}{0}

\subsection{Preliminaries}
\m{Preliminaries1}
 To present an arbitrary $D$-dimensional metric theory we consider the
Lagrangian:
 \be
 \lag_D = - \frac{1}{2\k_D}\lag_{g}(g_{\mu\nu}) + \lag_{m}(g_{\mu\nu},\Phi)\,.
 \m{lag-g}
 \ee
One includes derivatives up to the second order of the metric
$g_{\mu\nu}$ and $\Phi$, where the last defines matter sources
without concretization. Here and below ``hat'' means densities of
the wight +1, for example, $\hat g^{\mu\nu} = \sqrt{-g}g^{\mu\nu}$,
$\lag = \sqrt{-g}{\cal L}$, {\em etc}; $({,\alf}) \equiv \di_{\alf}$
means ordinary derivatives; Greek indexes enumerate $D$-dimensional
spacetime coordinates. Variation of (\ref{lag-g}) with respect to
$g^{\mu\nu}$ leads to the gravitational equations:
 \be
\hat{\cal G}_{\mu\nu} = \k_D\hat T_{\mu\nu}\, . \m{ddd+}
 \ee
Variation (\ref{lag-g}) with respect to $\Phi $ gives corresponding
matter equations. In this section, we derive the identities applying
both the N{\oe}ther theorem only and the N{\oe}ther theorem together
with the Belinfante procedure to the gravitational part of the
Lagrangian (\ref{lag-g}).

To examine perturbations, we need to consider the background
$D$-dimensional spacetime. Let it belong the metric $\Bar
g_{\mu\nu}$, from which the background Christoffel symbols $\Bar
\Gamma^\sig_{\tau\rho}$, covariant derivatives $\Bar D_\alf$ and the
background Riemannian tensor $\Bar R^\sig{}_{\tau\rho\pi}$ are
constructed; here and below ``bar'' means that a quantity is a
background one. We also use the background Lagrangian defined as
$\Bar \lag_D = \lag_D(\bar g_{\mu\nu},\Bar\Phi)$ and corresponding
background gravitational and matter equations. We assume that the
background fields $\bar g_{\mu\nu}$ and $\Bar\Phi$ satisfy the
background equations, and, thus are known (fixed). Here, for our
purposes, we incorporate the background metric $\Bar g_{\mu\nu}$
into $\lag_g$ changing the ordinary derivatives $\di_{\alf}$ by the
covariant $\BD_\alf$ ones in the usual way: $\di_\alf g_{\mu\nu}
\equiv \Bar D_\alf g_{\mu\nu} - \Bar \Gamma^\sig_{\alf\rho} \l.
g_{\mu\nu} \r|_\sig^\rho$. The generalized notation $\l. g_{\mu\nu}
\r|^\alf_\beta$ corresponds to the Lie derivative definition:
 \be
 \pounds_\xi g_{\mu\nu} =-\xi^\alf \BD_\alf g_{\mu\nu} + \l. g_{\mu\nu}
\r|^\alf_\beta \BD_\alf \xi^\beta\,, \m{LieQ}
 \ee
thus $\l. g_{\mu\nu} \r|^\alf_\beta=
-2g_{\beta(\mu}\delta^\alf_{\nu)}$; the choice of a sign corresponds
to \cite{Mitzk}. Then, using
 \bea
\Del^\alpha_{\mu\nu} & = &\Gamma^\alpha_{\mu\nu} - \Bar
{\Gamma}^\alpha_{\mu\nu} = \half g^{\alf\rho}\l( \BD_\mu g_{\rho\nu}
+ \BD_\nu g_{\rho\mu} - \BD_\rho g_{\mu\nu}\r)\, , \m{DeltaDefD}\\
  R^\lam{}_{\tau\rho\sig} & =&
\BD_\rho \Delta^\lam_{\tau\sig} -  \BD_\sig\Delta^\lam_{\tau\rho} +
 \Delta^\lam_{\rho\eta} \Delta^\eta_{\tau\sig} -
 \Delta^\lam_{\eta\sig} \Delta^\eta_{\tau\rho}
 + \Bar R^\lam{}_{\tau\rho\sig} = \delta R^\lam{}_{\tau\rho\sig} +
 \Bar R^\lam{}_{\tau\rho\sig}\,,
 \m{(17-DmD)}
 \eea
we transform the pure metric Lagrangian $\lag_g$ into an explicitly
covariant form: $\lag_g = \lag_c = {\lag}_c (g_{\mu\nu}; \Bar D_\alf
g_{\mu\nu}; \Bar D_{\beta\alf}  g_{\mu\nu})$, where $\Bar
D_{\beta\alf} \equiv\Bar D_\beta \Bar D_\alf $. Here and below
$\delta$ means a difference between a dynamical and a background
quantity. Thus for an arbitrary tensor density $Q$:
 \be \delta Q = Q- \Bar{ Q}
 \m{deltaQ}
 \ee
that is a finite (exact, not infinitesimal) perturbation.

\subsection{The N{\oe}ther method and identities}
\m{Nether1}

A direct application of the {\em canonical} N{\oe}ther procedure to
$-\lag_c/2\k_D $, as a {\em scalar density}, gives the identity $
-1/2\k_D \l({\Lix} \lag_c + \di_\alf(\xi^\alf \lag_c)\r) \equiv 0 $,
which is equivalent to
 \be - \Bar D_\alf \l[\hat u_\sig{}^\alf\xi^\sig + \hat
m_{\sig}{}^{\alf\tau}\Bar D_\tau \xi^\sig + \hat
n_\sig{}^{\alf\tau\beta}\Bar D_{\beta} \Bar D_{\tau}\xi^\sig\r]
\equiv \BD_\alf \hat \imath^\alf_C \equiv
 \di_\alf \hat \imath^\alf_C \equiv 0\, .
 \m{(+2+)}
 \ee
Here, the coefficients are defined in unique way (without
ambiguities) by the Lagrangian:
 \be
 \hat u_\sig{}^\alf =  -\frac{1}{\k_D}\l[\hat {\cal G}_\sig^\alf +
 \k_D \hat{\cal U}_{\sig}{}^\alf{}+\k_D \hat n_\lam{}^{\alf\tau\beta}\Bar
R^\lam_{~\tau\beta\sig}\r]\, , \m{(+3+)}
 \ee
 \be
  \hat m_\sig{}^{\alf\tau} = -\frac{1}{2\k_D}\l[
{{\delta \lag_c} \over {\delta (\Bar D_\alf g_{\mu\nu})}}
 \l.g_{\mu\nu}\r|^\tau_\sig -
 {{\di \lag_c} \over {\di (\Bar D_{\tau\alf} g_{\mu\nu})}}
\Bar D_\sig g_{\mu\nu} +
 {{\di \lag_c} \over {\di (\Bar D_{\beta\alf} g_{\mu\nu})}}
 \Bar D_\beta (\l.g_{\mu\nu}\r|^\tau_\sig)\r]\, ,
\m{(+4+)}
 \ee
 \be \hat n_\sig{}^{\alf\tau\beta} = -\frac{1}{4\k_D}
\l[{{\di \lag_c} \over {\di (\Bar D_\beta \Bar D_\alf g_{\mu\nu})}}
 \l.g_{\mu\nu}\r|^\tau_\sig +
 {{\di \lag_c} \over {\di (\Bar D_\tau \Bar D_\alf g_{\mu\nu})}}
 \l.g_{\mu\nu}\r|^\beta_\sig\r].
\m{(+5+)}
 \ee
It can be seen that $\hat n_\sig{}^{\alf\tau\beta} = \hat
n_\sig{}^{\alf\beta\tau}$. We use the notations
 \bea
\hat {\cal G}^\alf_\sig &\equiv & \frac{1}{2}{{\delta \lag_c} \over
{\delta g_{\mu\nu}}} \l.g_{\mu\nu}\r|^\alf_\sig \equiv - {{\delta
\lag_c} \over {\delta g_{\mu\alf}}} g_{\mu\sig} \equiv {{\delta
\lag_c} \over {\delta
g^{\mu\sig}}} g^{\mu\alf}\,,\m{calGmunu}\\
\hat{\cal U}_{\sig}{}^\alf{}
  & \equiv & -\frac{1}{2\k_D}\l( {{\di \lag_{c} } \over
{\di (\Bar D_{\beta\alf} g_{\mu\nu})}}\Bar D_{\sig\beta} g_{\mu\nu}
+{{\delta \lag_{c} } \over {\delta (\Bar D_\alf g_{\mu\nu})}}\Bar
D_\sig g_{\mu\nu} - \delta^\alf_\sig
\lag_{c} \r) \,, \m{TG}\\
{{\delta \lag_c} \over {\delta (\Bar D_\alf g_{\mu\nu})}}&\equiv
&{{\di \lag_c} \over {\di (\Bar D_\alf g_{\mu\nu})}} -
 \Bar D_\beta
\l({{\di \lag_c} \over {\di (\Bar D_{\alf\beta} g_{\mu\nu})}}\r)\,.
 \m{Lagrangian}
 \eea
 As usual, ${{\delta \lag_c}/{\delta g_{\mu\nu}}}$ means Lagrangian
derivatives, $\hat {\cal G}^\alf{}_\sig$ is exactly the symmetrical
left hand side of (\ref{ddd+}), and $\hat {\cal U}^\alf{}_\sig$ is
the generalized canonical energy-momentum related to the
gravitational Lagrangian in (\ref{lag-g}).

The generalized current in (\ref{(+2+)}) can be rewritten as
 \be
 \hat \imath^\alf_C =- \l[(\hat u_\sig{}^\alf + \hat n_\lam{}^{\alf\beta\gamma}
\Bar R^\lam_{~\beta\gamma\sig})\xi^\sig + \hat
 m^{\rho\alf\beta}\di_{[\beta}\xi_{\rho]} +
 \hat z^{\alf}_C\r]
 \m{(+7+)}
 \ee
where $z$-term  is defined as
 \be
 \hat z^{\alf}_C (\xi) =\hat
m^{\sig\alf\beta}\zeta_{\sig\beta}+ \hat n^{\rho\alf\beta\gamma}
\l(2 \BD_{\gamma}\zeta_{\beta\rho} - \BD_\rho
\zeta_{\beta\gamma}\r)\,,
 \m{(+8+)}
 \ee
and $2\zeta_{\rho\sigma} = - {\pounds}_\xi \Bg_{\rho\sigma} =
2\BD_{(\rho}\xi_{\sigma)}$. Thus, $z$-term disappears, if $\xi^\mu =
\lam^\mu$ is a Killing vector of the background spacetime. Then only
the current (\ref{(+7+)}) is determined by the energy-momentum $(u +
n\Bar R)$-term and the spin $m$-term.

Opening the identity (\ref{(+2+)}) and, since $\xi^\sig$, $
\di_{\alf}\xi^\sig$, $ \di_{\beta\alf}\xi^\sig$ and
$\di_{\gamma\beta\alf}
 \xi^\sig$
are arbitrary at every world point, equating independently to zero
the coefficients at $\xi^\sig$,  $\BD_{\alf} \xi^\sig$,
$\BD_{(\beta\alf)} \xi^\sig$ and $\BD_{(\gamma\beta\alf)} \xi^\sig$
we get a system of covariant identities of the Belinfante-Rosenfeld
type:
 \bea
 &{}& \BD_\alf  \hat u_\sig{}^\alf + \half
 \hat m_\lam{}^{\alf\rho} \Bar R^{~\lam}_{\sig~\rho\alf}
 +{\textstyle{1\over 3}} \hat n_\lam{}^{\alf\rho\gamma}
\BD_\gamma \Bar R^{~\lam}_{\sig~\rho\alf}
  \equiv 0, \m{(+9+1)}\\
&{}&    \hat u_\sig{}^\alf + \BD_\lam \hat m_{\sig}{}^{\lam \alf} +
\hat n_\lam{}^{\tau\alf\rho}
 \Bar R^{~\lam}_{\sig~\rho\tau} +{\textstyle{2\over 3}} \hat
 n_{\sig}{}^{\lam\tau\rho}\Bar R^{\alf}_{~\tau\rho\lam} \equiv 0,
 \m{(+9+2)}\\ &{}&
 \hat m_{\sig}{}^{(\alf\beta)}+
\BD_\lam  \hat n_{\sig}{}^{\lam(\alf\beta)} \equiv 0, \m{(+9+3)}\\
&{}&
 \hat
 n^{(\alf\beta\gamma)}_\sig \equiv 0.
 \m{(+9+4)}
 \eea
These are the generalization to arbitrary curved backgrounds of the
expressions given by Mitzkevich \cite{Mitzk}. The above identities are
not independent: the first one (\ref{(+9+1)}) is a consequence of
the other three.

As a rule, a divergence $\lag'= \hat d^\nu{}_{,\nu}$ in the
Lagrangian is not essential for deriving field equations. However,
in a definition of canonical conserved quantities it plays an
important role. Let us illustrate this below. For the scalar density $\lag'=
\hat d^\nu{}_{,\nu}$ one has the N{\oe}ther identity
 \be ({\Lix} \hat
d^\alf + \xi^\alf \hat d^\nu{}_{,\nu})_{,\alf} \equiv 0\,,
 \m{(+2'+)}
 \ee
which has to be considered together with (\ref{(+2+)}). An
expression under divergence in (\ref{(+2'+)}) defines an additional
contribution
 \bea
\hat \imath'^\alf &=&-\l[ \hat u'_\sig{}^\alf\xi^\sig + \hat
m'_\sig{}^{\alf\tau}\Bar D_\tau \xi^\sig \r]\,; \m{di}
\\
\hat u'_\sig{}^\alf & =& 2\BD_\beta (\delta^{[\alf}_\sig \hat
d^{\beta]} )\,,~~~~ \hat m'_\sig{}^{\alf\beta}=
 2 \delta^{[\alf}_\sig \hat d^{\beta]}\,,~~~~
\hat n'_\sig{}^{\alf\beta\gamma} =0\,.\m{umn}
 \eea
into the current (\ref{(+7+)}). Note that the construction of the
quantities (\ref{umn}) in the additional current (\ref{di}) does not
depend on the inner structure of $\hat d^{\nu}$.

\subsection{The Belinfante symmetrization}
\m{Belinfante1}

Using the Belinfante rule \cite{Belinfante} generalized in \cite{PK}
we define the Belinfante correction:
 \be
\hat s^{\alf\beta\sig} =- \hat s^{\beta\alf\sig} = -
 \hat m_\lam{}^{\sig[\alf} \bar g^{\beta]\lam} -
 \hat m_\lam{}^{\alf[\sig} \bar g^{\beta]\lam} + \hat m_\lam{}^{\beta[\sig}
 \bar g^{\alf]\lam}\, ,
 \m{(+13+)}
 \ee
and modify (\ref{(+7+)}). Thus, the Belinfante corrected current is
 \be \hat \imath^\alf_B = \hat \imath^\alf_C
+\BD_{\beta}(\hat s^{\alf\beta\sig}\xi_{\sig})= -\l[\l( \hat
u_\sig{}^{\alf} +\hat n_\lam{}^{\alf\beta\gamma}\Bar
R^\lam_{~\beta\gamma\sig} - \BD_{\beta} \hat
s^{\alf\beta}{}_{\sig}\r)  \xi^\sig+
 \hat z^{\alf}_B(\xi)\r]\,.
 \m{B-current}
 \ee
By definition, it does not contain the spin term (coefficient at
$\di_{[\beta}\xi_{\rho]}$). The new $z$-term
 \be
 \hat z^{\alf}_B(\xi) =
\l(\Bar g^{\lam\sig}\hat
 m_\lam{}^{\beta\alf}+ 2\Bar g^{\lam[\sig} \hat m_\lam{}^{\alf]\beta}
\r)
 \zeta_{\sig\beta} +
 \hat
n_{\rho}{}^{\alf\beta\gamma} \l(2 \BD_{\gamma}\zeta_{\beta}^{\rho} -
\BD^\rho \zeta_{\beta\gamma}\r)\,
 \m{Zmu-2}
 \ee
disappears for Killing vectors of the background as well. Due to
antisymmetry in (\ref{(+13+)}), this current (\ref{B-current}) is
also identically conserved:
 \be
 \di_\alf\hat \imath^\alf_B \equiv\BD_{\alf}  \hat \imath^{\alf}_B
 \equiv 0\,.
 \m{B-CL}
 \ee
It is important to  note that the Belinfante procedure cancels the
addition (\ref{di}) induced by a divergence in Lagrangian.

Because the currents $\hat \imath^\alf_C $ and $\hat \imath^\alf_B $
satisfy the identities (\ref{(+2+)}) and (\ref{B-CL}) they have to
be expressed through correspondent antisymmetrical tensor densities
(superpotentials) $\hat \imath^{\alf\beta}_C $ and $\hat
\imath^{\alf\beta}_B $, for which $\di_\alf\di_\beta\hat
\imath^{\alf\beta}_C \equiv\di_\alf\di_\beta\hat
\imath^{\alf\beta}_B \equiv 0$. Indeed, following the standard
prescription \cite{Petrov2008} and (\ref{(+9+2)}) - (\ref{(+9+4)}),
one can construct these superpotentials satisfying
 \bea
 \hat \imath^\alf_C &\equiv &\BD_{\beta} \hat \imath^{\alf\beta}_C \equiv
\di_{\beta}   \hat \imath^{\alf\beta}_C\,,
 \m{(+10+)}\\
 \hat \imath^\alf_B &\equiv &\BD_{\beta} \hat \imath^{\alf\beta}_B \equiv
\di_{\beta}   \hat \imath^{\alf\beta}_B\,.
 \m{Bsup}
 \eea
These demonstrate a principal form of a connection of the currents
constructed here with the superpotentials in
 \cite{Petrov2009}.

Here, the coefficients (\ref{(+3+)}) - (\ref{(+5+)}) are {\it
uniquely} defined by the pure metric part of the Lagrangian
(\ref{lag-g}). Consequently, N{\oe}ther's and N{\oe}ther-Belinfante's
procedures give uniquely defined currents. The same claim is
related, of course, to all the quantities constructed below for
perturbations and based on the identities presented here.

\section{Currents in arbitrary $D$-dimensional
metric theories}
 \m{CurrentsInD-dimensions}
 \setcounter{equation}{0}

In previous section, we have derived the identities and the
identically conserved currents related to the {\em external}
background spacetime which looks as an {\em auxiliary} structure.
Here, we use these results to describe perturbations, which are
determined when one (dynamical) solution of the theory is
considered as a perturbed system with respect to another solution
(background) of the same theory. Perturbations in such a scenario
are exact (not infinitesimal or approximate), and the background
spacetime acquires a {\em real} sense, and is not just an auxiliary structure.
The same scheme, which can be named as bimetric,
has been explored in \cite{Petrov2009}. Linear and higher order approximations simply
follow once the exact form is presented.

\subsection{Canonical N{\oe}ther current}
 \m{NCS}

The expressions presented in subsection \ref{Nether1} are maximally
adopted to construct N{\oe}ther canonical conserved quantities in
the framework of the bimetric formulation. Following the
Katz-Bi\v c\'ak-Lynden-Bell (KBL) ideology \cite{KBL} we construct
the Lagrangian:
 \be
 \lag_{G} = -\frac{1}{2\k_D}\l(\lag_{g} - \Bar\lag_{g} + \di_\alf \hat d^\alf\r)\, .
 \m{ArbitraryLagPert}
 \ee
This Lagrangian, constructed for perturbations, has to be vanishing
for vanishing perturbations. Thus usually $\hat d^\alf$ is chosen to
satisfy this requirement, i.e., to disappear for vanishing
perturbations, see, e.g., \cite{KBL,DerKatzOgushi}. Applying the
barred procedure to (\ref{(+7+)}) and taking into account the
divergence (using (\ref{di}) and (\ref{umn})), one obtains the
current corresponding to (\ref{ArbitraryLagPert}):
 $
\hat I^\alf_C = \hat \imath^\alf_C - \Bar{\hat \imath^\alf_C} +\hat
\imath'^\alf
 $.
We then use the dynamical equations (\ref{ddd+}) in $\hat
u_{\sig}{}^{\alf}$. We change ${\cal G_{\mu\nu}}$ (as a part of
$\hat u_{\sig}{}^{\alf}$, see (\ref{(+3+)})) by the matter
energy-momentum ${T_{\mu\nu}}$ at right hand side of (\ref{ddd+}).
Next, we do the same combining $\Bar{\hat u}_{\sig}{}^{\alf}$ and
the barred equations (\ref{ddd+}). In the result one obtains that
the identically conserved current $\hat I^\alf_C$ related to
(\ref{ArbitraryLagPert}) transforms into the current
 \be
\hat {\cal I}^\alf_{C}(\xi) = {}_C\hat {\Theta}_{\sig}{}^\alf
\xi^\sig + {}_C\hat {\cal M}^{\sig\alf\beta}\di_{[\sig}\xi_{\beta]}
+ {}_C\hat {\cal Z}^\alf(\xi)\,
  \m{(+7+A)}
 \ee
for perturbations. Now, the conservation law:
 \be
\Bar D_\alf \hat {\cal I}^\alf_{C}(\xi)= \di_\alf \hat {\cal
I}^\alf_{C}(\xi) = 0\,
 \m{identity-Ic}
 \ee
holds in place due to the field equations (not identically). The
generalized canonical energy-momentum, spin and $Z$-term are
 \bea
{}_C\hat {\Theta}_{\sig}{}^\alf &\equiv &  \delta\hat T^\alf_\sig +
\delta \hat{\cal U}_{\sig}{}^\alf +
\k^{-1}_D\BD_\beta(\delta^{[\alf}_\sig\hat d^{\beta]}) \,,
\m{7A'}\\
{}_{C}\hat {\cal M}^{\sig\alf\beta} &\equiv & \delta\hat
 m_{\rho}{}^{\alf\beta}\bar g^{\sig\rho}- \k^{-1}_D\bar g^{\sig[\alf}\hat
 d^{\beta]}\,,
 \m{7A''}\\
{}_{C}\hat {\cal Z}^\alf(\xi) &\equiv &
  -\delta \hat z^{\alf}_{C} + \k^{-1}_D\zeta^{[\alf}_\beta\hat d^{\beta]}
\,
 \m{7A'''}
 \eea
 where perturbations are constructed  following the general definition
(\ref{deltaQ}). To calculate perturbations one has to use quantities
presented in (\ref{ddd+}), (\ref{(+4+)}), (\ref{(+5+)}), (\ref{TG})
and (\ref{(+8+)}).

\subsection{Belinfante symmetrized current}
 \m{BCS}

To construct the Belinfante corrected conserved currents for the
perturbed system (\ref{ArbitraryLagPert}) we turn to subsection
\ref{Belinfante1}.  We subtract the barred expression
(\ref{B-current}) from the original one (\ref{B-current}):
  $
 \hat I^\alf_{B}= \hat \imath^\alf_{B} - \Bar{\hat \imath^\alf_{B}}
  $.
Of course, the same is obtained after applying the
N{\oe}ther-Belinfante method directly to the Lagrangian in
(\ref{ArbitraryLagPert}). Again, using the equations (\ref{ddd+})
and their barred version in $\hat u_{\sig}{}^{\alf}$ and $\Bar{\hat
u}_{\sig}{}^{\alf}$, the current $\hat I^\alf_B$ related to
(\ref{ArbitraryLagPert}) transforms into
 \be
\hat {\cal I}^\alf_{B}(\xi) = {}_B\hat{\Theta}_{\sig}{}^\alf\xi^\sig
+ {}_B\hat {\cal Z}^\alf(\xi) \, .
 \m{(+15+A)}
 \ee
Thus, one has a conservation law
 \be
\Bar D_\alf \hat {\cal I}^\alf_{B}(\xi)= \di_\alf \hat {\cal
I}^\alf_{B}(\xi) = 0\,
 \m{identity-Ib}
 \ee
for perturbations satisfying the field equations. As it has to be,
the current (\ref{(+15+A)}) does not contain a spin term, unlike
(\ref{(+7+A)}). The Belinfante corrected energy-momentum and
$Z$-term are
 \bea
{}_B\hat{\Theta}_{\sig}{}^\alf &\equiv & \delta\hat T^\alf_\sig +
\delta \hat{\cal U}_{\sig}{}^\alf  + \BD_{\beta}\delta \hat
s^{\alf\beta}{}_{\sig} \, ,
\m{7A+}\\
{}_{B}\hat {\cal Z}^\alf(\xi) &\equiv & -\delta\hat
z^{\alf}_{B}(\xi)\, .
 \m{(7A++)}
 \eea
To calculate perturbations one has to use quantities presented in
(\ref{ddd+}), (\ref{(+4+)}), (\ref{(+5+)}), (\ref{TG}),
(\ref{(+13+)}) and (\ref{Zmu-2}).

Note that, the energy-momenta (\ref{7A'}) and (\ref{7A+}) are
separated into the two parts: matter and pure gravitational ones.
However, this separation is conventional because the relation
between these two parts can be changed easily by another combination
with the field equations (\ref{ddd+}) that is quite permissible.

\section{Currents in the
Einstein-Gauss-Bonnet gravity}
 \m{EGBcurrents}
\setcounter{equation}{0}

\subsection{Preliminaries}
\m{Preliminaries2}

In this section, we apply the theoretical results of the previous
sections to derive the explicit structure of the currents in the EGB
gravity of both the kinds (\ref{(+7+A)}) and (\ref{(+15+A)}). We
search for the expressions in the most general form: they are not to
be restricted by any concrete backgrounds or dynamic solutions. As a
basis for calculation, we use expressions presented in the Appendix.
Because Z-terms in (\ref{(+7+A)}) and (\ref{(+15+A)}) disappear for
the Killing vectors of the background they are not so essential.
Therefore, we do not give their explicit form, since these can be
easily reconstructed using the auxiliary expressions from the
Appendix.

The action of the Einstein $D$-dimensional theory with a bare
cosmological term $\Lambda_0$ and a Gauss-Bonnet correction term
(see, for example, \cite{DT2}) is
 \bea
 S  &=& -\frac{1}{2\k_D}\int d^D x\lag_{EGB} +\int d^D x\lag_{m}
\nonumber\\&=& -\frac{1}{2\k_D}\int d^D x \sqrt{-g} \l[R -
2\Lambda_0 +
 \alpha(RR)_{GB}\r]  +\int d^D x\lag_{m}\,,
 \m{EGBaction}
 \eea
 \be
(RR)_{GB} \equiv  R_{\alf\beta\gamma\delta}^2
 - 4 R_{\alf\beta}^2+ R^2 \,,
 \m{GB-term}
 \ee
where $\k_D = 2\Omega_{D-2}G_D> 0$  and $\alpha >0$; $G_D$ is the
$D$-dimension Newton's constant. Below, the subscripts ``${}_{E}$''
is related to the pure Einstein part of the action
(\ref{EGBaction}), and the subscript ``${}_{GB}$'' is related to the
Gauss-Bonnet part connected with $\alf$-coefficient. The field
equations that follow from (\ref{EGBaction}) have the form of
(\ref{ddd+}) with
 \bea
 \hat {\cal G}^{\mu\nu} &=& - \frac{\delta}{\delta
g_{\mu\nu}}\lag_{EGB}
  = {\sqrt{-g}}\l\{\l(R^{\mu\nu} - \half g^{\mu\nu}
 R + g^{\mu\nu}\Lambda_0\r)\r.\nonumber \\
 &+& \l.2\alpha\l[RR^{\mu\nu} -
 2 R^{\mu}{}_{\sig}{}^\nu{}_{\rho} R^{\sig\rho} +
 R^{\mu}{}_{\sig\rho\tau}R^{\nu\sig\rho\tau} - 2 R^{\mu}{}_{\sig}
 R^{\sig\nu}  -{\textstyle \frac{1}{4}} g^{\mu\nu}
 \l(RR\r)_{GB}\r] \r\}.
 \m{EGBequationsE}
 \eea

 \subsection{Canonical prescription}
 \m{C-currentEGB}

Let us turn to the current (\ref{(+7+A)}). Its structure (\ref{7A'})
- (\ref{7A'''}) essentially depends on the divergence in the
Lagrangian. We choose the divergence induced by the Katz-Lifshits
approach \cite{KatzLivshits} (see discussions in
\cite{Petrov2009,Petrov2009a}). Thus, in (\ref{7A'}) - (\ref{7A'''})
we consider
 \be
\hat d^\alf =  {}_{(E)}\hat d^\alf + {}_{(GB)}\hat d^\alf = {2}
\Delta^{[\tau}_{\tau\beta}\hat g^{\alf]\beta}  +
 4 \alf\l(
\hat R_\rho{}^{\beta\tau\alf} - 2\hat R^{[\tau}_\rho g^{\alf]\beta}-
2\delta^{[\tau}_\rho \hat R^{\alf]\beta}+ \delta^{[\tau}_\rho
g^{\alf]\beta} \hat R\r)\Delta^\rho_{\tau\beta}\, .
 \m{divd}
 \ee
In $D$-dimensional GR, the Katz and Livshits superpotential
\cite{KatzLivshits} turns out {\em uniquely} the KBL superpotential
\cite{KBL}. In EGB gravity, their superpotential (essentially
connected with (\ref{divd}) and the GB term (\ref{GB-term}))
naturally transfers into the KBL superpotential for $D=4$. Thus,
although the GB term does not affect the derivation of the field
equations for $D=4$, it plays an important role (as a criterium) in
definition of superpotentials of canonical type. The use of the term
(\ref{GB-term}) in the Lagrangian even in {\em four dimensions}
turns out important when the other ideas are elaborated. For
example, in \cite{Olea} Olea includes the GB term to regularize
conserved quantities, in \cite{M-Olea} Mi\v{s}kovi\'{c} and Olea
show that the standard holographic regularization procedure of AdS
gravity with counterterms is topological and, thus, can be presented
by the addition of the GB term.

Now, not calculating $Z$-term, we construct (\ref{7A'}) and
(\ref{7A''}). The matter part in (\ref{7A'}) is defined by the
sources in (\ref{ddd+}), however its concrete form will not be presented here.
For present purpose, it is more interesting to focus on the
gravitational part of (\ref{7A'}) denoted below as ${}_C\!\hat {\cal
T}_\sig{}^\alf$. Thus, we calculate (\ref{7A'}) with the use of
(\ref{TG-EGB}) and (\ref{divd}):
 \bea
 {}_C\!\hat {\Theta}_\sig{}^\alf &=& \delta\hat {T}_\sig{}^\alf
 + {}_C\!\hat {\cal T}_\sig{}^\alf = \sqrt{-g} {T}_\sig{}^\alf
 -\sqrt{-\Bar g}\,\, \Bar{T}_\sig{}^\alf+
\frac{1}{2\k_D}\delta^\alf_\sig\l[\Bar R_{\rho\tau}\hat l^{\rho\tau}
-
2\Lambda_0\delta\sqrt{-g} \r]\nonumber\\
 &+& \frac{\sqrt{-g}}{\k_D} \l[
\l(\Delta^\alf_{\rho[\tau}\Delta^\pi_{\pi]\sig}  +
\Delta^\alf_{\rho[\sig}\Delta^\pi_{\pi]\tau}+ \delta^\alf_\sig
\Delta^\pi_{\beta[\tau}\Delta^\beta_{\pi]\rho} - \delta^\alf_\rho
\Delta^\beta_{\beta[\sig}\Delta^\pi_{\pi]\tau}\r) g^{\tau\rho} \r] \nonumber\\
&+& \frac{1 }{\k_D}\BD_\beta(\delta^{[\alf}_\sig{}_{(GB)}\hat
d^{\beta]}) + \frac{\alf }{2\k_D}\delta^\alf_\sig
\,\delta\!\l( \hat RR \r)_{GB} \nonumber\\
&+& \frac{2\alf\sqrt{-g} }{\k_D} \l[\l(R^{\alf\beta\rho}{}_\tau -4
g^{\rho[\alf}R^{\beta]}_\tau + Rg^{\rho[\alf} \delta^{\beta]}_\tau
\r)\Bar D_\sig\Delta^{\tau}_{\beta\rho}\r.\nonumber\\
&+&\l. 2g^{\beta\mu}\l(\BD_\beta R^{\alf\nu} +
2\Delta^{(\alf}_{\beta\rho}R^{\nu]\rho} \r)\Delta^{\tau}_{\sig(\mu}
g_{\nu)\tau} -g^{\rho(\alf}\Delta^{\nu)}_{\sig\rho}\di_\nu R\r] \,.
 \m{canonical-EM}
 \eea
The symbol $\delta$ without subscripts, once again, means a perturbation of a
quantity with respect to a background (\ref{deltaQ}); we also use
the notation: $\hat l^{\rho\tau} =\delta\hat
g^{\rho\tau}=\delta(\sqrt{-g} g^{\rho\tau})  $ \cite{PK}.

To calculate the spin term (\ref{7A''}) we use (\ref{mEGB})
subtracting the barred (\ref{mEGB}) and taking into account
(\ref{divd}). Thus
 \bea
 {}_C\hat{\cal M}^{\sig\alf\beta} &= &
-\frac{\sqrt{- g}}{2\k_D}\l[ \Delta^{\tau}_{\rho\tau}\l(2\Bar
g^{\sig[\alf} g^{\beta]\rho} + \Bar g^{\sig\rho}g^{\alf\beta}\r) -
\Delta^{\alf}_{\rho\tau}\l(2\Bar g^{\sig[\tau} g^{\beta]\rho} + \Bar
g^{\sig\rho}g^{\tau\beta}\r) \r]
\nonumber\\
&-& \frac{1}{\k_D}\bar g^{\sig[\alf}{}_{(GB)}\hat d^{\beta]} +
\frac{2\alf\sqrt{-g}}{\k_D}
 \l[ R^{\alf\tau\rho}{}_\lam \Delta^{\beta}_{\tau\rho}
-2R^{\alf(\tau\beta)}{}_\rho\Delta^{\rho}_{\tau\lam} \r]\Bar
g^{\lam\sig}
 \nonumber\\
&+ & \frac{4\alf\sqrt{-g}}{\k_D}\l[4 g^{\rho[\alf}
R^{\beta]}_{\tau}\Delta^\tau_{\rho\lam} + 2 R^{[\alf}_\lam
g^{\tau]\rho}\Delta^{\beta}_{\tau\rho}+ 2 g^{\alf[\beta}
R^{\tau]}_{\rho}\Delta^{\rho}_{\tau\lam}+g^{\tau\beta}\l(
R^{\alf}_\rho \Delta^{\rho}_{\tau\lam}- R^{\rho}_{(\tau}
\Delta^{\alf}_{\lam)\rho}\r)\r.
 \nonumber\\&-& \l. \l(g^{\rho\beta}\Delta^\tau_{\rho(\tau}+
 g^{\rho\tau}\Delta^\beta_{\rho(\tau}-g^{\tau\beta}\Delta^\rho_{\rho(\tau}\r)
  R^{\alf}_{\lam)}
 \r]\Bar g^{\lam\sig} -\frac{4\alf}{\k_D}\Bar g^{\lam\sig}
\BD_{(\tau}\delta\l( \hat g^{\tau\beta}R^\alf_{\lam)} \r)
 \nonumber\\ &+&
 \frac{2\alf\sqrt{- g}}{\k_D}R\l[\Delta^{[\alf}_{\rho\lam} g^{\rho]\beta}+
\Delta^{(\alf}_{\rho\lam} g^{\beta)\rho)} \r]\Bar
g^{\lam\sig}+\frac{2\alf}{\k_D}\Bar g^{\sig(\alf}
\BD_{\rho}\,\delta\l( \hat g^{\rho)\beta}R \r)\,.\m{CALmEGB}
 \eea
As expected, these expressions disappear for vanishing perturbations.
The Einstein parts in (\ref{canonical-EM}) and
(\ref{CALmEGB}) exactly coincide with the energy-momentum and the
spin tensor presented in \cite{KBL}. We do not present explicitly
the  terms with ${}_{(GB)}\hat d^\alf$ because this does not
simplify the expression as a whole.

\subsection{Prescription of the generalized Belinfante procedure}
 \m{B-currentEGB}

Here, we turn to the current (\ref{(+15+A)}). Its structure, see
(\ref{7A+}) and (\ref{(7A++)}), unlike the canonical case, does not
depend on a spin term and a divergence in the Lagrangian. As before,
we do not consider $Z$- term. Constructing (\ref{7A+}) we calculate
explicitly the pure gravitational part only, it is denoted below as
${}_B{\hat{\cal T}^{\sig\alf}}$. We substitute (\ref{TG-EGB}) and
(\ref{s-EGB}) into (\ref{7A+}), raise indices,  use the field
equations (\ref{ddd+}) with (\ref{EGBequationsE}) and, as a result,
obtain
 \bea {}_B\hat\Theta^{\sig\alf} & = & \delta \hat T^{(\alf}_\rho\Bar
g^{\sig)\rho} +{}_B{\hat{\cal T}^{\sig\alf}} = \sqrt{-g}\,
T^{(\alf}_\rho\Bar g^{\sig)\rho} -\sqrt{-\Bar g}\,\,\Bar
T^{\alf\sig} \nonumber\\&+& \frac{1}{2\k_D}\l[\hat l^{\rho\tau} \Bar
R_{\rho\tau} \Bar g^{\alf\sig}+ 2{\hat l}^{\lam[\alf} \bar
R^{\sig]}_\lam -2\Bar g^{\alf\sig}\Lambda_0
\delta\sqrt{-g}\r]\nonumber
\\ & + & \frac{1}{2\k_D}\l[ \l(\hat l^{\alf\sig}\Bar
g^{\rho\tau} - \Bar g^{\alf\sig}\hat l^{\rho\tau}\r) \Bar D_\tau
\Delta^\lambda_{\rho\lambda} + 2\l(\hat l^{\rho\tau} \Bar
g^{\lambda(\alf} - \Bar g^{\rho\tau}\hat l^{\lambda(\alf}\r) \Bar
D_\tau \Delta^{\sig)}_{\lambda\rho}\r]
 \nonumber \\ &+ & \frac{1}{2\k_D}\l[ \Bar
g^{\rho\tau}\l(\hat g^{\alf\sig}\Delta^\lambda_{\rho\lambda}
\Delta^\eta_{\tau\eta}+ 2\hat
g^{\lambda\eta}\Delta^{(\alf}_{\lambda\rho}\Delta^{\sig)}_{\eta\tau}\r)
+ \hat g^{\lambda\eta} \Bar
g^{\alf\sig}\Delta^\tau_{\rho\lambda}\Delta^\rho_{\tau\eta}\r]\nonumber
\\ & + & \frac{1}{\k_D}\l[\Bar
g^{\rho\tau}\l(\Delta^{\lambda}_{\tau\eta}\Delta^{(\alf}_{\lambda\rho}\hat
g^{\sig)\eta} -
2\Delta^{\lambda}_{\tau\lambda}\Delta^{(\alf}_{\eta\rho}\hat
g^{\sig)\eta}\r)+ \hat g^{\lambda\eta}
\l(\Delta^{\tau}_{\rho\tau}\Delta^{(\alf}_{\lambda\eta} -
\Delta^{\tau}_{\lambda\eta}\Delta^{(\alf}_{\rho\tau} -
\Delta^{\tau}_{\lambda\rho}\Delta^{(\alf}_{\eta\tau}\r)\Bar
g^{\sig)\rho} \r]\nonumber\\
&+& \frac{2\alf\sqrt{- g}}{\k_D} \Bar
g^{\lam\sig}\l[\l(R^{\alf\beta\rho}{}_\tau -4
g^{\rho[\alf}R^{\beta]}_\tau + Rg^{\rho[\alf} \delta^{\beta]}_\tau
\r)\Bar D_\lam\Delta^{\tau}_{\beta\rho}- g^{\rho(\alf}
\Delta^{\beta)}_{\lam\rho}\BD_\beta R\r.
 \nonumber\\&+&\l. 2g^{\beta\mu}\l(\BD_\beta
R^{\alf\nu}+
 2R^{\rho(\alf}\Delta^{\nu)}_{\beta\rho}\r)
 \Delta^{\tau}_{\lam(\mu}g_{\nu)\tau} \r] +
 \frac{\alf}{2\k_D}\Bar g^{\alf\sig}\delta\,(\hat RR)_{GB}
 +
\Bar D_\beta \delta {}_{(GB)}\hat s^{\alf\beta\sig}\nonumber\\&-&
\frac{\alf\sqrt{-g}}{\k_D}\Bar g^{\lam[\alf}
\l[R^{\sig]}{}_{\pi\rho\tau}R_{\lam}{}^{\pi\rho\tau}- 2
R^{\sig]}{}_{\pi\lam\rho}R^{\pi\rho} -
2R^{\sig]}_{\rho}R^{\rho}_\lam + R^{\sig]}_{\lam}R\r]\,.
 \m{B-Theta-EGB} \eea
The Einstein part exactly coincides with the one presented in
\cite{PK}. Recall that, even this part (symmetrized) is not
symmetrical in general, see \cite{PK}. Here, we do not open the
divergence of the GB-part $\delta {}_{(GB)}\hat s^{\alf\beta\sig}$
of the Belinfante correction, see (\ref{(+13+)}), because this does
not simplify the expression; for calculations it is more convenient
to use the already known/calculated components obtained with using
(\ref{s-EGB}). The {\em symmetrical} matter part and the last line
in (\ref{B-Theta-EGB}) are the result of the secondary use of the
field equations (\ref{ddd+}) with (\ref{EGBequationsE}). Like
(\ref{canonical-EM}), the energy-momentum (\ref{B-Theta-EGB})
disappears for vanishing perturbations, note that the barred last
line in (\ref{B-Theta-EGB}) vanishes due to the antisymmetrization.

\section{Applications}
 \m{Applications}
 \setcounter{equation}{0}

 \subsection{Weak flat gravitational waves}
 \m{waves}

 Here, we use formulae from previous
 sections to calculate energy density for weak flat gravitational
 wave in the EGB gravity. Such a gravitational wave propagates in
 $D$-dimensional flat spacetime and is described by the linearized
vacuum equations (\ref{ddd+}) with (\ref{EGBequationsE}). Due to the
requirement of the flat background the GB part in
(\ref{EGBequationsE}), being quadratic in curvature components, does
not contribute to the linearized equations, and $\Lambda_0=0$. Thus,
effectively, these are the linear Einstein equations in $D$
dimensions without cosmological term. Assume that
 $g_{\mu\nu}= \eta_{\mu\nu}+h_{\mu\nu}$
and that the Lorentz coordinates are used. Then one has
$\eta_{\mu\nu}= {\rm diag}(-1,\,1\ldots 1)$ and $\Delta^*_{**}\sim
h_{**,*}$. Linear Einstein equations after applying the standard
technique of the $TT$-gauge \cite{MTW} have a form
 \be
 R_{\mu\nu}=-\half  h_{\mu\nu,\alf}{}^{,\alf} =0\,.
 \m{linearEGB}
 \ee
Assuming $h_{\mu\nu} = h_{\mu\nu}(t-x)= h_{\mu\nu}(x^0-x^1)$ one
obtains $h_{0\alf}=h_{1\alf}=h_\alf^\alf=0$, and non-zero components
are $h_{kl}$ where the Latin indices from the middle of alphabet
numerate: $k,l\ldots=2,3\ldots,D-1$.

Let us turn to the canonical prescription. To calculate the energy
density one has to calculate the 0-component of the current
(\ref{(+7+A)}) with the Killing vector $ \xi^\alf = \lambda^\alf =
(-1,\,{\bf 0});~\lambda_\alf = (1,\,{\bf 0})\,. $ Then, only the
00-component of the pure gravitational energy-momentum ${}_C\hat
{\cal T}_0{}^0$ in (\ref{canonical-EM}) contributes (without spin
term (\ref{CALmEGB})); for the linearized wave we calculate
${}_C\hat {\cal T}_0{}^0$ up to the second order. In direct
calculations we take into account: a) a flat background with zero
Riemannian tensor and its contractions; b) zero linear parts of the
Ricci tensor and curvature scalar due to (\ref{linearEGB}); c)
proportionality $\Delta^*_{**}\sim h_{**,*}$. In the end we show
that, including the quadratic terms, the GB part of ${}_C\hat {\cal
T}_0{}^0$ is equal to zero as a whole. Considering the Einstein part
one obtains in quadratic approximation:
 \be
 {}_C{\cal T}_0{}^0 = - \frac{1}{4\k_D}\sum^{D-1}_{k,l=2}\dot h^2_{kl}
 \m{canonical-00}
  \ee
where dot means differentiation with respect to $t=x^0$.

Now we turn to the Belinfante prescription to calculate the
0-component of the current (\ref{(+15+A)}). Keeping in mind the
above assumptions and using the Killing vector $\lambda^\alf$ we
need to calculate only the 00-component of the pure gravitational
energy momentum ${}_B\hat {\cal T}^{00}$ in (\ref{B-Theta-EGB}).
Again the GB part is equal to zero in quadratic approximation. Thus,
${}_B\hat {\cal T}^{00}$ is also defined by the Einstein part only:
 \be
 {}_B{\cal T}^{00} =  \frac{1}{4\k_D}\sum^{D-1}_{k,l=2}\dot h^2_{kl}\,.
 \m{Belinfante-00}
  \ee

Contracting both (\ref{canonical-00}) and (\ref{Belinfante-00}) with
the Killing vector $\lambda^\alf$ one obtains the unique expression
for the energy density of the flat weak gravitational waves:
 \be
 {\cal I}^{0}_C = {\cal I}^{0}_B = \frac{1}{4\k_D}\sum^{D-1}_{k,l=2}\dot h^2_{kl}\,.
 \m{E-density}
  \ee
This is in a full correspondence with the standard results in 4D GR
\cite{MTW}. Because the equations (\ref{linearEGB}), in fact, are
the Einstein ones the energy density (\ref{E-density}) is
acceptable. Thus, the new energy-momentum expressions applied here
to describe flat gravitational waves satisfy simple, but important
and non-trivial, test. Indeed, the equations (\ref{linearEGB}) are from
the {\em start} the linearized EGB equations, not the Einstein
equations; the currents (\ref{(+7+A)}) (with (\ref{canonical-EM})
and (\ref{CALmEGB})) and (\ref{(+15+A)}) (with (\ref{B-Theta-EGB}))
from the {\em start} have been constructed in the framework of the
EGB gravity, not the Einstein gravity.

\subsection{Radiative 3D black hole of the Kaluza-Klein type}
\m{radiative}

In this subsection, we apply the new formulae to describe mass
fluxes for interesting and important solutions obtained recently in
the works \cite{MaedaDadhich2,MaedaDadhich3}. The main assumption is
that a spacetime is to be locally homeomorphic to ${\cal M}^d
\times{\cal K}^{D-d} $ with the metric $g_{\mu\nu} ={\diag}
(g_{AB},r^2_0\gamma_{ab})$, $A,B = 0,\cdots,d-1;~a,b=d,\cdots,D-1$.
Thus, $g_{AB}$ is an arbitrary Lorentz metric on ${\cal M}^d$;
$\gamma_{ab}$ is the unit metric on the $(D-d)$-dimensional space of
constant curvature ${\cal K}^{D-d}$ with $k=0,\,\pm 1$. Factor $r_0$
is a small scale of extra dimensions. Vacuum gravitational equations
${\cal G}^\mu{}_\nu =0$ (see (\ref{EGBequationsE})) are decomposed
into two separate systems ${\cal G}^A{}_B =0$ and ${\cal G}^a{}_b
=0$. The first one is a tensorial equation on ${\cal M}^d$, whereas
the second one is a constraint for it. However, to obtain more
interesting solutions one has to consider a special case, when the
expression ${\cal G}^A{}_B$ disappears {\em identically}. This is
possible for $d \le 4$ only. In this case, constants are chosen so
as to suppress the coefficients in ${\cal G}^A{}_B$, which is
possible when $D\ge d+2$, $k=-1$ and $\Lambda_0 <0$. After taking
into account all of the above, a single governing equation is ${\cal
G}^a{}_b =0$. In reality it is the unique scalar equation on ${\cal
M}^d$ because ${\cal G}^a{}_b \sim \delta^a_b$ and depends on
$g_{AB}$ only.

Here, we consider the case $D=6$ and $d=3$ presented in
\cite{MaedaDadhich3}. A suitable set of constraints for the
constants is $r^2_0 = 12\alf =-3/\Lambda_0$. Then, the unique scalar
equation is
 \be
{}_{(d)}\!{R}=2\Lambda_0\,,
 \m{R=2L}
 \ee
where subscript `${}_{(d)}$' imply that a quantity is constructed
with the use of $g_{AB}$ only. This scalar equation is satisfied by
both the static and the radiative metric \cite{MaedaDadhich3}. Here,
 for constructing the mass fluxes, it is quite appropriate to use
the new current expressions. We apply them to radiative solution
$g_{AB}(v,r)$ of the Vaidya type \cite{MaedaDadhich3}:
 \be
 ds^2 = - fdv^2 +2dvdr +r^2d\phi\,,\qquad f\equiv r^2/l^2  + q(v)/r -\mu(v)\,
 \m{KK-vmetric}
 \ee
where $l^2 \equiv -3/\Lambda_0$. In this conctrete case $\mu(v)$ and
$q(v)$ depend on the advanced time $v$. Non-zero components
corresponding to the solution (\ref{KK-vmetric}), $d=3$ sector, are
as follows. Metric components are
 $
 g_{00} = -f(v), \,g_{01} = 1, \,g_{22} = r^2\, ;
 $
 Christoffel symbols, components of Riemannian
and Ricci tensors, curvature scalar, and components of the Einstein tensor
are
 \bea
 \Gamma^1_{00} &=& (f{f}' - \dot f)/2 ,\,
 \Gamma^0_{00} = {{f}'}/{2} ,\,
 \Gamma^1_{01} = -{{f}'}/{2} ,\,
 \Gamma^2_{12} =  1/r ,\,
 \Gamma^1_{22} = -rf,\,
 \Gamma^0_{22} = -r,
 \m{Christ-v}\\
R^{0101}&=& \half f''\,, ~~~R^{0212}= -\frac{f'}{2r^3}\,
,~~~R^{1212}= -\frac{1}{2r^3}(ff' +\dot f)\, ;\nonumber\\
R^{11}& =& - \frac{1}{2r}\l[f\l(rf'' + {f'}\r) + \dot f\r]\, ,~~~
~~~R^{01} = - \frac{1}{2r}\l(rf'' + {f'}\r)\, ,~~~ R^{22} =
-\frac{f'}{r^3}\, ; \nonumber\\ R &=& - \frac{1}{r}\l(rf'' +
2{f'}\r)\, ,  \m{R-v}\\
G^0_0 &=& G^1_1 =1/l^2
 -q/2r^3,~~G^1_0 = (\dot \mu r-\dot q)/2r^2,~~ G^2_2 = 1/l^2
 +q/r^3\,,\m{ET-v}
 \eea
where `prime' and `dot' mean $\di/\di r$ and $\di/\di v$. The scalar
curvature of $D-d=3$ sector is
 \be
{}_{(D-d)}\!{R}= 6k/r_0^2=2\Lambda_0 = -1/2\alpha\,.
 \m{(D-d)R=2L}
 \ee

In fact, (\ref{KK-vmetric}) - (\ref{ET-v}) together with
(\ref{(D-d)R=2L}) present 6D solution in EGB gravity. Whereas
(\ref{KK-vmetric}) - (\ref{ET-v}) without (\ref{(D-d)R=2L}) can be
considered as a solution to the Einstein 3D equations on ${\cal
M}^3$, which are not vacuum equations with redefined cosmological
constant $\Lambda = -1/l^2$:
 \be
 {}_{(3)}\!{R}_{AB}-\half g_{AB}{}_{(3)}\!{R} +g_{AB}\Lambda =
\k_3 {T}_{AB}\,.
 \m{EEwithTAU}
 \ee
A natural treating in \cite{MaedaDadhich3} is that ${T}_{AB}$,
corresponding to (\ref{ET-v}), is created by extra dimensions.

Here, both the full 6D presentation in the framework of the EGB
gravity and the 3D interpretation (\ref{EEwithTAU}) are explored. We
consider a cylinder $S := r=\rm const$. The wall $S$ can be thought
as 5D timelike hypersurface in 6D spacetime, or as 2D timelike
hypersurface in 3D spacetime; $\di \Sigma$ is an intersection of $S$
with a lightlike hypersurface $v=$ const. To present a mass flux
through $\di \Sigma$ one has to calculate the component $\hat {\cal
I}^1$ of the currents (\ref{(+7+A)}) or (\ref{(+15+A)}) and
integrate it over $\di \Sigma$. The total background metric in the
6D derivation can be chosen as $\Bar g_{\mu\nu}= \Bar g_{AB} \times
r^2_0\gamma_{ab}$ with the AdS${}_3$ metric $\Bar g_{AB}$ presented
by $\Bar f \equiv r^2/l^2 + 1$ in the element of the type
(\ref{KK-vmetric}), see \cite{Petrov2009a}. Whereas the background
metric in the 3D derivation is chosen as the same AdS${}_3$ metric
$\Bar g_{AB}$ only. Background components are derived from
(\ref{KK-vmetric}) - (\ref{(D-d)R=2L}) after applying the barred
procedure. To calculate the {\em mass} flux we use the {\em
timelike} background Killing vector $
 \xi^\alf = \lam^\alf =(-1, {\bf 0});~\lam_\alf =(\Bar f,\,-1, {\bf 0})
$ where ${\bf 0}$ includes  all the rest space dimensions both for
6D and for 3D derivations.

Let us present results of calculations in the framework of the
canonical prescription of subsection \ref{NCS} with the formulae of
subsection \ref{C-currentEGB} in detail. Turn to the 6D derivation.
Then using all the components (\ref{KK-vmetric}) - (\ref{(D-d)R=2L})
together with the barred ones,  we substitute them into
(\ref{canonical-EM}) and (\ref{CALmEGB}). Recall that Z($\lam$)-term
disappears, and note that, unlike subsection \ref{waves}, we {\em
need} to calculate the spin term (\ref{CALmEGB}). After very
prolonged and cumbersome calculations we obtain for (\ref{(+7+A)}):
${}_{(E)}\hat {\cal I}^1_C(\lam) \equiv{}_{(GB)}\hat {\cal
I}^1_C(\lam) \equiv 0$ that gives $\hat {\cal I}^1_C(\lam) \equiv
0$.  Thus,
 \be
 \dot M= \oint_{\di
\Sigma}dx^{D-2}\hat{\cal I}^1_C \equiv 0\,.
 \m{flux=0}
 \ee

At a first glance, the result (\ref{flux=0}) looks strange.
However, it is in full correspondence with the results in
\cite{Petrov2009a} where we have just calculated the masses of $d=3$
objects in $D = 6$ EGB gravity with the use of the superpotentials
constructed in \cite{Petrov2009}. Let us demonstrate this
correspondence. A general expression for the total mass has been
obtained as a surface integral in $D=6$ dimensions
\cite{Petrov2009a}:
  \be
M =  \oint_{\di\Sigma} dx^{D-2}\,\sqrt{-\Bar g_D}\, {\cal I}^{01} =
\oint_{r\goto\infty} d\phi\sqrt{-\Bar g_d}\, {\cal
I}^{01}\,\oint_{r_0} dx^{D-d}\sqrt{-\Bar g_{D-d}}=
V_{r_0}\oint_{r\goto\infty} d\phi\sqrt{-\Bar g_d}\, {\cal I}^{01}.
 \m{charges=0}
 \ee
Integration over $d=3$ sector gives zero. In the canonical approach,
it is so because ${\cal I}^{01}_{C}\equiv 0$ (in more
details
 ${}_{(GB)}\hat {\cal I}^{01}_{C}\equiv
-{}_{(E)}\hat {\cal I}^{01}_{C}\neq 0$). For the Belinfante
corrected approach the integration over $d=3$ sector gives zero due
to the asymptotic behaviour of ${\cal I}^{01}_B$, in spite of ${\cal
I}^{01}_B\neq 0$. The formula (\ref{charges=0}) shows that one needs
to consider two possibilities: (i) when extra $D-d=3$ dimensions are
not compactified; (ii) when they are compactified by appropriate
identifications.

In the case (i), one has to consider objects as 6 dimensional ones.
Of course, in spite of  $V_{r_0} \goto \infty$, their masses in
(\ref{charges=0}) have to be equated to zero. Next, because $M$ is
defined for arbitrary $\di \Sigma$ one has $M_{\di \Sigma_0} =
M_{\di \Sigma_ 1} =0$ that determines null flux through $\di
\Sigma$. One finds just a correspondence of (\ref{flux=0}) with the corresponding
conclusion in \cite{Petrov2009a}.

The case (ii), in our opinion, has a more physical sense. Now,
$V_{r_0}$ in (\ref{charges=0}) is finite. Then, because ${\cal
I}^{01} \sim 1/\k_6$ one can set $\k_3 = \k_6/V_{r_0}$, and, really,
in (\ref{charges=0}) one has $M\sim 1/\k_3$.  This means that the 6D
Einstein constant $\k_6$ is reduced to the 3-dimensional one $\k_3$
that is the standard Kaluza-Klein prescription. One has to reject 6D
derivation and turn to 3D derivation with the Einstein presentation
(\ref{EEwithTAU}) and with the evident interpretation of $\k_3$. In
this case, null mass is quite unacceptable. Therefore one has to use
ingredients of the Einstein theory only, and not the EGB one. Thus, applying
superpotentials constructed in \cite{Petrov2009}, we have used their
Einstein parts only, changing $\k_6$ by $\k_3 $: ${}_{(E)}\hat {\cal
I}^{01}= (\mu +1 - q/r)/2\k_3 $ that gives acceptable mass for the
solution (\ref{KK-vmetric}) on the AdS${}_3$ background
\cite{Petrov2009a}:
 \be
 {}_{(E)}M= \oint_{\di
\Sigma}dx^{D-2}{}_{(E)}\hat{\cal I}^{01}_C = (\mu +1)\pi/\k_3\,.
 \m{Msuper}
 \ee

Exploring the expressions for currents presented here one has to use the
Einstein interpretation (\ref{EEwithTAU}) also. However now, unlike
the superpotential application, we cannot use the reduced the
Einstein part of the current of the 6D description because
${}_{(GB)}\hat {\cal I}^{1}_C\equiv {}_{(E)}\hat {\cal I}^{1}_C
\equiv 0$. Nevertheless, there is no a contradiction. Recall that in
6D picture we are based on the {\em vacuum} EGB equation (pure
gravitational), whereas for the 3D description (\ref{EEwithTAU})
constructing currents (see (\ref{(+7+A)}) and (\ref{7A'})) we must
use the created matter at the right hand side of (\ref{EEwithTAU}).
Thus, in 3D derivation, the component $T^1{}_0= (\dot\mu r -\dot q)/
2\k_3r^2 $ in (\ref{EEwithTAU}) just determines ${}_{(E)}\hat {\cal
I}^{1}_C=\sqrt{-g}T^1{}_0\lam^0= -(\dot\mu -\dot q/r)/2\k_3 $ by a
crucial way. This gives the flux:
 \be
{}_{(E)}\dot M = \oint_{\di \Sigma}dx^{D-2}{}_{(E)}\hat{\cal I}^1_C
=-\dot\mu\pi/\k_3\,.
 \m{fluxNOT=0}
 \ee
Differentiating mass (\ref{Msuper}) (obtained in the framework of
the superpotential derivation) with respect to $v$ one obtains:
${}_{(E)}\dot M= \dot\mu\pi/\k_3 $ \cite{Petrov2009a}. One can see a
difference in a sign, however there is no contradiction. A
simplified differentiation of $M$ with respect to $v$ gives, in
fact, an absolute value of the flux. A check with using $\hat{\cal
I}^{1}_Ñ = \di_0 \hat{\cal I}^{10}_Ñ$ and antisymmetry $\hat{\cal
I}^{10}_Ñ = -\hat{\cal I}^{01}_Ñ$ shows a correspondence in signs
also.

The same conclusions follow when the Belinfante symmetrization
method developed in subsections \ref{BCS} and \ref{B-currentEGB} is
applied. Though, unlike the canonical approach, in 6D description:
${\cal I}^{1}_B\neq 0$; and, in 3D interpretation, ${}_{(E)}{\cal
I}^{1}_B$ is not determined by the created energy-momentum in
(\ref{EEwithTAU}) only. However, due to the asymptotic behaviour,
additional terms do not contribute into the final expressions after
integration. Thus, once again, in the case (i), one obtains a zero mass
flux (\ref{flux=0}), whereas, in the case (ii), one needs to use the
3D (\ref{EEwithTAU}) interpretation and has the flux
(\ref{fluxNOT=0}).

\section{Concluding remarks}
 \m{Concluding}
  \setcounter{equation}{0}
 The modern development of multidimensional metric theories themselves, naturally,
 includes/induces a development of methods for constructing
 conserved quantities.
In the present paper, in the framework of the $D$-dimensional EGB
gravity we have presented the explicit covariant expressions for the
conserved currents of perturbations of an arbitrary type on
arbitrary curved backgrounds. The two methods, {\em canonical} and
{\em Belinfante corrected}, have been applied. The main parts in the
structure of the canonical and Belinfante corrected currents, which
are the energy-momentum tensors (\ref{canonical-EM}) and
(\ref{B-Theta-EGB}), are the generalization of the Einstein
pseudotensor \cite{Einstein16} and of the Papapetrou pseudotensor
\cite{Papapetrou48}, respectively.

Together with an evident academic interest, a construction of such
currents can be very useful in applications. Indeed, many solutions
of modified metric theories need to be examined in detail. It is
necessary because frequently such solutions look quite exotic, and one
has to understand the physical meaning they represent, how contradictive
or non-contradictive they are, {\em etc}. Thus, by presenting rules for
constructing conserved quantities including conserved currents that are
important physical characteristics of objects, we present the
instrument for analyzing these objects.

Applications in section \ref{Applications}, should be seen as tests for the
new expressions. Indeed, weak flat gravitational wave is the
standard object with well studied properties; also properties of the 3D
radiating black holes already have been studied by us in
\cite{Petrov2009a}. However, the direct calculation of the mass flux
using the new current expressions can be viewed as an
important independent result. In \cite{MaedaDadhich2,MaedaDadhich3},
the matter presented by the energy-momentum at the right hand side
of (\ref{EEwithTAU}) is treated as being created by all the extra
dimensions as a whole. Of course, such a derivation differs from the
standard Kaluza-Klein picture where each of compactified extra
dimensions determines its own charge. Nevetherless, as we show in
subsection \ref{radiative}, the compactified dimensions are reduced
in the standard Kaluza-Klein prescription. Also, we demonstrate that
the created matter in (\ref{EEwithTAU}) determines the {\em
classically} defined mass (\ref{Msuper}) (see \cite{Petrov2009a})
and mass flux (\ref{fluxNOT=0}) of the objects. Thus, keeping in
mind the above comments, we support the claim of the authors of
\cite{MaedaDadhich2,MaedaDadhich3} that their solutions present the
objects of the Kaluza-Klein type.

It is important to compare our results with the results by Cai, Cao
and Ohta \cite{Cai+}. The authors, in the framework of the Lovelock gravity
of an arbitrary order, have constructed and analyzed a new
solutions analogous to (\ref{KK-vmetric})-(\ref{(D-d)R=2L}), only
static. Using the Wald technique \cite{Wald}, they have proved that
the objects corresponding to such solutions have zero entropy and,
consequently, zero mass. This coincides with our conclusions.
Indeed, \cite{Cai+} uses the EGB Lagrangian in all D
dimensions. In our consideration, in the case (i), when objects are
examined  in all 6 dimensions, both mass of the objects
\cite{Petrov2009a} and their mass flux (even in radiative regime
(\ref{flux=0})) are equal to zero.

In the future, we intend to continue to construct conserved quantities in EGB gravity,
and present conserved currents in the framework of the {\em
symmetrical} approach (see Introduction). Also, we plan to use the
new expressions, both for the superpotentials and for the currents
in EGB gravity, to describe interesting solutions, say, 4D objects
in $D$-dimensional EGB gravity \cite{MaedaDadhich2}.

Lastly, the possibility of a connection between AdS gravity and a
conformal field theory (CFT) living on its boundary induces a
considerable attention. A definition of conserved quantities and an
existence of nonzero energy for asymptotically AdS vacuum spacetime
could be useful to identify the AdS/CFT correspondence at the
boundary. To define finite conserved quantities one requires a
regularization procedure, the mechanism of which does not invoke the
substraction of background configurations. In this context, one of
more popular approaches is the {\em boundary counterterm method}. It
is developing more intensively in the framework of  EGB gravity, one
can recommend, e.g., interesting works \cite{Olea2,Radu1} and
numerous references there in. Unlike the boundary counterterm
method, the prescriptions explored here (and in
\cite{Petrov2009,Petrov2008,Petrov2009a}) use the background
spacetime in a crucial way. Nevertheless, it could be very useful to
compare these approaches, and we plan to do this in future.

\subsection*{Acknowledgments} The author is very grateful
to Rodrigo Olea and Rong-Gen Cai for explanations of their works,
 fruitful discussions and useful recommendations. The author also
 thanks very much Deepak Baskaran for a significant improvement of
 English.

%Katz for explanation his work with Lifshits and useful
%conversations. The work is supported by the grant No.
%09-02-01315-a of the Russian Foundation for Basic Research.

\appendix

\section{Auxiliary expressions in EGB gravity}
 \m{GEinEGB}
 \setcounter{equation}{0}

In this appendix, we calculate the coefficients, which are derived
following the definitions (\ref{(+3+)}), (\ref{(+4+)}) and
(\ref{(+5+)}), and correspond to the Lagrangian $\lag_{EGB}$ in
(\ref{EGBaction}). However, at the first it is useful to present the
next derivatives:
 \bea
&{}& \frac{-2\k_D}{\sqrt{-g}}\frac{\di \lag_{EGB}}{\di g_{\mu\nu}} =
 \frac{-2\k_D}{\sqrt{-g}}\l(\frac{\di \lag_{E}}{\di g_{\mu\nu}}+
 \frac{\di \lag_{GB}}{\di g_{\mu\nu}} \r)\nonumber\\& =& \half
 g^{\mu\nu}(R - 2\Lambda_0)  + \l[
 2\l(g^{\rho(\mu}\Delta^{\nu)}_{\alf[\sig}\Delta^{\alf}_{\rho]\tau}
 +  g^{\alf\rho}\Delta^{(\mu}_{\tau[\sig}\Delta^{\nu)}_{\rho]\alf}
+g^{\rho(\mu}\Bar D_{[\sig}\Delta^{\nu)}_{\rho]\tau}\r)g^{\tau\sig}-
R^{\mu\nu}\r]\nonumber\\&{+}& \frac{\alf
g^{\mu\nu}}{2}\l(R_{\lam\tau\rho\sig}R^{\lam\tau\rho\sig} - 4
R_{\rho\sig}R^{\rho\sig} +R^2\r)\nonumber\\
&+&2\alf\l[2\l(g^{\lam(\mu}\Delta^{\nu)}_{\alf\sig}\Delta^{\alf}_{\rho\tau}
+g^{\alf\lam}\Delta^{(\mu}_{\tau\sig}\Delta^{\nu)}_{\rho\alf}+
g^{\lam(\mu}\Bar D_\sig\Delta^{\nu)}_{\rho\tau}\r)
R_\lam{}^{\tau\rho\sig}- R^{\mu\tau\rho\sig}R^\nu{}_{\tau\rho\sig}
\r]  \nonumber\\&{-}& 8\alf\l[
 2\l(g^{\rho(\mu}\Delta^{\nu)}_{\alf[\sig}\Delta^{\alf}_{\rho]\tau}
 +  g^{\alf\rho}\Delta^{(\mu}_{\tau[\sig}\Delta^{\nu)}_{\rho]\alf}
+g^{\rho(\mu}\Bar D_{[\sig}\Delta^{\nu)}_{\rho]\tau}\r)R^{\tau\sig}-
R^{\mu\rho}R^{\nu}_\rho\r]\nonumber\\&{+}& 2\alf R\l[
 2\l(g^{\rho(\mu}\Delta^{\nu)}_{\alf[\sig}\Delta^{\alf}_{\rho]\tau}
 +  g^{\alf\rho}\Delta^{(\mu}_{\tau[\sig}\Delta^{\nu)}_{\rho]\alf}
+g^{\rho(\mu}\Bar D_{[\sig}\Delta^{\nu)}_{\rho]\tau}\r)g^{\tau\sig}-
R^{\mu\nu}\r]\,;
  \m{DiLgmn}
 \eea
 \bea
&{}& \frac{-2\k_D}{\sqrt{-g}}\frac{\di \lag_{EGB}}{\di(\Bar D_\alf
g_{\mu\nu})} =
 \frac{-2\k_D}{\sqrt{-g}}\l(\frac{\di \lag_{E}}{\di(\Bar D_\alf
g_{\mu\nu})}+
 \frac{\di \lag_{GB}}{\di(\Bar D_\alf
g_{\mu\nu})} \r)\nonumber\\& =& 2
\l[\Delta^{\alf}_{\sig\rho}g^{\sig[\rho}g^{\mu]\nu} +
g^{\alf\sig}\Delta^{(\mu}_{\sig\rho}g^{\nu)\rho} -
g^{\alf(\mu}\Delta^{\nu)}_{\sig\rho}g^{\sig\rho} \r]\nonumber\\& +&
4\alf\l[2R^{\alf\sig\rho(\mu}\Delta^{\nu)}_{\sig\rho}
-\Delta^{\alf}_{\sig\rho}R^{\sig\mu\nu\rho}\r]\nonumber\\& -&
4\alf\l[2 R^{\alf\sig}\Delta^{(\mu}_{\sig\rho}g^{\nu)\rho} -
2g^{\alf(\mu}\Delta^{\nu)}_{\sig\rho}R^{\sig\rho} + 2
g^{\alf\sig}\Delta^{(\mu}_{\sig\rho}R^{\nu)\rho}
-2R^{\alf(\mu}\Delta^{\nu)}_{\sig\rho}g^{\sig\rho}\r.\nonumber\\ &+&
\l. \Delta^{\alf}_{\sig\rho}R^{\sig\rho}g^{\mu\nu}+
\Delta^{\alf}_{\sig\rho}g^{\sig\rho}R^{\mu\nu} -
2\Delta^{\alf}_{\sig\rho}R^{\sig(\mu}g^{\nu)\rho}\r]
 \nonumber\\& +&
 4\alf R
\l[\Delta^{\alf}_{\sig\rho}g^{\sig[\rho}g^{\mu]\nu} +
g^{\alf\sig}\Delta^{(\mu}_{\sig\rho}g^{\nu)\rho} -
g^{\alf(\mu}\Delta^{\nu)}_{\sig\rho}g^{\sig\rho} \r]\,;
 \m{DiLDgmn}
 \eea
 \bea
&{}& \frac{-2\k_D}{\sqrt{-g}}\frac{\di \lag_{EGB}}{\di(\Bar
D_{\beta\alf} g_{\mu\nu})} =
 \frac{-2\k_D}{\sqrt{-g}}\l(\frac{\di \lag_{E}}{\di(\Bar D_{\beta\alf}
g_{\mu\nu})}+
 \frac{\di \lag_{GB}}{\di(\Bar D_{\beta\alf}
g_{\mu\nu})} \r)\nonumber\\& =& \l[g^{\alf(\mu}g^{\nu)\beta}-
g^{\alf\beta}g^{\mu\nu}\r]\nonumber\\& +&
2\alf\l[2R^{\alf(\mu\nu)\beta} - 4 R^{\alf(\mu}g^{\nu)\beta} +
2g^{\mu\nu}R^{\alf\beta} + 2g^{\alf\beta}R^{\mu\nu} + R
\l(g^{\alf(\mu}g^{\nu)\beta}- g^{\alf\beta}g^{\mu\nu} \r)\r]\,.
  \m{DiLDDgmn}
 \eea

With using (\ref{DiLgmn}) - (\ref{DiLDDgmn}) we calculate the
coefficients (\ref{(+3+)}) - (\ref{(+5+)}) for the EGB Lagrangian in
(\ref{EGBaction}). Because (\ref{(+3+)}) is defined by
(\ref{calGmunu}), (\ref{TG}) and (\ref{(+5+)}) we present these
parts separately. Thus
  \bea
 \hat {\cal G}^{\alf}_{\sig} &=& {}_{(E)}\hat {\cal G}^{\alf}_{\sig}
 +{}_{(GB)}\hat {\cal G}^{\alf}_{\sig}\m{calGmunuEGB}\\
&=&{\sqrt{-g}}\l(R^{\alf}_{\sig} -\half \delta^{\alf}_{\sig}
 R + \delta^{\alf}_{\sig}\Lambda_0\r) \nonumber\\
 &+& \alpha{\sqrt{-g}}\l[2\l(RR^{\alf}_{\sig} -
 2 R^{\alf}{}_{\tau\sig\rho} R^{\tau\rho} +
 R^{\alf}{}_{\pi\rho\tau}R_{\sig}{}^{\pi\rho\tau} - 2 R^{\alf}_{\rho}
 R^{\rho\sig}\r)  -\half\delta^{\alf}_{\sig}
 (RR)_{GB} \r].
 \nonumber
 \eea
 \bea
 \hat {\cal U}_{\sig}{}^{\alf} &=& {}_{(E)}\hat {\cal U}_{\sig}{}^{\alf}
 +{}_{(GB)}\hat {\cal U}_{\sig}{}^{\alf}\m{TG-EGB}\\&=&
-\frac{\sqrt{- g}}{2\k_D}\l[2g^{\rho[\tau} \Bar
D_\sig\Delta^{\alf]}_{\rho\tau}- \delta^\alf_\sig(R-2\Lambda_0) \r]\nonumber\\
&+& \frac{2\alf\sqrt{- g}}{\k_D} \l[\l(R^{\alf\beta\rho}{}_\tau -4
g^{\rho[\alf}R^{\beta]}_\tau + Rg^{\rho[\alf} \delta^{\beta]}_\tau
\r)\Bar D_\sig\Delta^{\tau}_{\beta\rho}\r.
 \nonumber\\&+&\l. 2g^{\beta\mu}\l(\BD_\beta
R^{\alf\nu}+
 2R^{\rho(\alf}\Delta^{\nu)}_{\beta\rho}\r)
 \Delta^{\tau}_{\sig(\mu}g_{\nu)\tau} - g^{\rho(\alf}
\Delta^{\beta)}_{\sig\rho}\BD_\beta R \r]+ \frac{\alf\sqrt{-
g}}{2\k_D}\delta^\alf_\sig(RR)_{GB}\,.\nonumber
 \eea
 \bea
 {\hat m}_{\sig}{}^{\alf\beta} &= & {}_{(E)}{\hat m}_{\sig}{}^{\alf\beta}
 +{}_{(GB)}{\hat m}_{\sig}{}^{\alf\beta}
 \m{mEGB}\\ &=&
-\frac{\sqrt{- g}}{2\k_D}\l[\delta^\alf_\sig
\Delta^{\beta}_{\rho\tau}g^{\rho\tau} - 2
\Delta^{\alf}_{\sig\rho}g^{\beta\rho} + \Delta^{\rho}_{\rho\sig}
g^{\alf\beta}  \r]
 \nonumber\\ &+ & \frac{2\alf\sqrt{-g}}{\k_D}
 \l[\ R^{\alf\tau\rho}{}_\sig \Delta^{\beta}_{\tau\rho}
-2R^{\alf(\tau\beta)}{}_\rho\Delta^{\rho}_{\tau\sig}
\r] \nonumber\\
&+ & \frac{4\alf\sqrt{-g}}{\k_D}\l[4 g^{\rho[\alf}
R^{\beta]}_{\tau}\Delta^\tau_{\rho\sig} + 2 R^{[\alf}_\sig
g^{\tau]\rho}\Delta^{\beta}_{\tau\rho}+ 2 g^{\alf[\beta}
R^{\tau]}_{\rho}\Delta^{\rho}_{\tau\sig} \r.
 \nonumber\\&-& \l. g^{\tau\beta}\l(\BD_{(\tau} R^\alf_{\sig)}
  + R^{\rho}_{(\tau}
\Delta^{\alf}_{\sig)\rho} - R^{\alf}_\rho
\Delta^{\rho}_{\tau\sig}\r) \r]
 \nonumber\\&-&
 \frac{\alf\sqrt{- g}}{\k_D}\l[\l(\delta^\alf_\sig
\Delta^{\beta}_{\rho\tau}g^{\rho\tau} - 2
\Delta^{\alf}_{\sig\rho}g^{\beta\rho} + \Delta^{\rho}_{\rho\sig}
g^{\alf\beta}\r)R - 2\delta^{(\alf}_\sig g^{\tau)\beta} \di_\tau R
\r]\nonumber
 \eea
 \bea
 {\hat n}_{\sig}{}^{\lam\alf\beta} &= &{}_{(E)}{\hat
 n}_{\sig}{}^{\lam\alf\beta} + {}_{(GB)}{\hat n}_{\sig}{}^{\lam\alf\beta}
\m{nEGB} \\ &=& \frac{\sqrt{-
g}}{2\k_D}\l\{g^{\alf\beta}\delta^\lam_\sig
-g^{\lam(\alf}\delta^{\beta)}_\sig \r\}
 \nonumber\\ &+& \frac{\alf\sqrt{-g}}{\k_D}
 \l\{-2R_\sig{}^{(\alf\beta)\lam} - 4 R_\sig^{\lam}g^{\alf\beta} +
 4 R_\sig^{(\alf}g^{\beta)\lam} + R\l(g^{\alf\beta}\delta^\lam_\sig
-g^{\lam(\alf}\delta^{\beta)}_\sig\r)\r\}\, .
 \nonumber
 \eea
It was checked directly that the coefficients $\hat u_\sig{}^\alf$
in (\ref{(+3+)}) (calculated with the use of (\ref{calGmunuEGB}),
(\ref{TG-EGB}) and (\ref{nEGB})), and the coefficients (\ref{mEGB})
and (\ref{nEGB}) themselves satisfy exactly the identities
(\ref{(+9+1)}) - (\ref{(+9+4)}).

At last, using (\ref{mEGB}) we calculate the Belinfante correction
(\ref{(+13+)}) for the EGB gravity:
 \bea
 \hat s^{\alf\beta\sig} &=& {}_{(E)}\hat s^{\alf\beta\sig} +
  {}_{(GB)}\hat s^{\alf\beta\sig}
 \nonumber \\&=&
  \frac{\sqrt{-g}}{\k_D}
\l[\Delta^{[\alf}_{\tau\rho}\Bar g^{\beta]\sig}g^{\tau\rho}+
\Delta^{\rho}_{\lam\rho}g^{\sig[\alf}\Bar g^{\beta]\lam}
-\Delta^{\sig}_{\lam\rho}g^{\rho[\alf}\Bar g^{\beta]\lam} -
2\Delta^{[\alf}_{\lam\rho}\Bar g^{\beta]\lam}g^{\rho\sig}+
\Delta^{[\alf}_{\lam\rho}g^{\beta]\rho}\Bar g^{\sig\lam}
\r]\nonumber\\
&+& \frac{2\alf\sqrt{-g}}{\k_D}\l[ \Bar
g^{\lam[\alf}\l(R^{\beta]\sig\tau}{}_\rho -
2R^{\beta]\tau\sig}{}_\rho \r)\Delta^\rho_{\lam\tau} +\Bar
g^{\lam[\alf}\Delta^{\beta]}_{\tau\rho}R^{\sig\tau\rho}{}_\lam +\Bar
g^{\lam[\alf}R^{\beta]\tau\rho}{}_\lam\Delta^{\sig}_{\tau\rho}\r.
\nonumber\\ &+& \l. \Bar
g^{\sig\lam}\l(R_\lam{}^{\tau\rho[\alf}\Delta^{\beta]}_{\tau\rho} -
\textstyle{\frac{3}{2}}R^{\alf\beta\tau}{}_\rho
\Delta^\rho_{\lam\tau}\r)\r]\nonumber\\ &+&
\frac{4\alf\sqrt{-g}}{\k_D}\l[\l(R^\tau_\lam
g^{\rho(\sig}\Delta^{\alf)}_{\tau\rho}- g^{\tau\rho}R^{(\sig}_\lam
\Delta^{\alf)}_{\tau\rho}- g^{\sig\alf}R^{\tau}_\rho
\Delta^{\rho}_{\tau\lam}  + \BD_{(\tau}R^{(\sig}_{\lam)}
g^{\alf)\tau}+ R^{\rho}_{(\tau}\Delta^{(\sig}_{\lam)\rho}
g^{\alf)\tau}\r)\Bar g^{\beta\lam}\r. \nonumber\\
&-& \l(R^\tau_\lam g^{\rho(\sig}\Delta^{\beta)}_{\tau\rho}-
g^{\tau\rho}R^{(\sig}_\lam \Delta^{\beta)}_{\tau\rho}-
g^{\sig\beta}R^{\tau}_\rho \Delta^{\rho}_{\tau\lam}  +
\BD_{(\tau}R^{(\sig}_{\lam)} g^{\beta)\tau}+
R^{\rho}_{(\tau}\Delta^{(\sig}_{\lam)\rho}
g^{\beta)\tau}\r)\Bar g^{\alf\lam} \nonumber\\
&+& \l.\l(2g^{\tau[\alf}R^{\beta]}_\rho \Delta^{\rho}_{\tau\lam}
-R^\tau_\lam g^{\rho[\alf}\Delta^{\beta]}_{\tau\rho}+
g^{\tau\rho}R^{[\alf}_\lam \Delta^{\beta]}_{\tau\rho}-
\BD_{(\tau}R^{[\alf}_{\lam)} g^{\beta]\tau}-
R^{\rho}_{(\tau}\Delta^{[\alf}_{\lam)\rho} g^{\beta]\tau}\r)\Bar
g^{\sig\lam}\r]
 \nonumber\\
  &+& \frac{2\alf\sqrt{-g}}{\k_D}\l[\l( g^{\sig[\alf}\Bar
g^{\beta]\lam}\Delta^\rho_{\lam\rho}-\Delta^{(\sig}_{\lam\rho}g^{\alf)\rho}\Bar
g^{\beta\lam}+ \Delta^{(\sig}_{\lam\rho}g^{\beta)\rho}\Bar
g^{\alf\lam} - \Bar
g^{\sig[\alf}\Delta^{\beta]}_{\lam\rho}g^{\lam\rho}- \Bar
g^{\sig\lam} g^{\rho[\alf}\Delta^{\beta]}_{\lam\rho}\r)R\r.\nonumber\\
&+&\l. \l(\Bar g^{\lam[\alf}g^{\beta]\sig} + \Bar g^{\sig[\alf}
g^{\beta]\lam} \r)\di_\lam R \r]
 \m{s-EGB}
 \eea
Of course, the Einstein part exactly coincides with the one
presented in \cite{PK}.

\ed